\documentclass[useAMS,usenatbib]{mn2e} 
\usepackage{color}
\usepackage{psfig}
\usepackage{epsfig}
\usepackage{amsbsy}  
\usepackage{amssymb}

\newcommand{\be}{\begin{equation}} 
\newcommand{\ee}{\end{equation}}
\newcommand{\bea}{\begin{eqnarray}} 
\newcommand{\eea}{\end{eqnarray}}

\newcommand{\no}{\noindent} 
\newcommand{\COSMOMC}{{\sc CosmoMC}}
\newcommand{\CAMB}{{\sc camb}}

\newcommand*{\unit}[1]{\ensuremath{\mathrm{\, #1}}}
\newcommand*{\erg}{\unit{erg}}
\newcommand*{\cm}{\unit{cm}}
\newcommand*{\second}{\unit{s}}
\newcommand*{\E}[1]{\ensuremath{\times 10^{#1}}}
\newcommand*{\keV}{\unit{keV}}

\def\lt{<}

\def\cm{{\rm\thinspace cm}}
\def\erg{{\rm\thinspace erg}}

\def\keV{{\rm\thinspace keV}}

\def\Mpc{{\rm\thinspace Mpc}}

\def\spose#1{\hbox to 0pt{#1\hss}}
\def\approxpropto{\mathrel{\spose{\lower 3pt\hbox{$\sim$}}
	\raise 2.0pt\hbox{$\propto$}}}
 \def \lleq {\lower0.9ex\hbox{$\buildrel \lt \over \sim$} ~}

\title[Constraints on modified gravity from XLF data] 
{Constraints on modified gravity from the observed X-ray luminosity function of galaxy clusters}
\author[D.~Rapetti, S.~W.~Allen, A.~Mantz \& H.~Ebeling] {David~Rapetti${}^{1}$\thanks{Email: drapetti@slac.stanford.edu},
  Steven~W.~Allen${}^{1}$, Adam~Mantz${}^{1}$ and Harald~Ebeling$^2$\\
  ${}^1$ Kavli Institute for Particle Astrophysics and Cosmology at\\
  Stanford University, 452 Lomita Mall, Stanford 94305-4085, CA, USA, and\\
  SLAC National Accelerator Laboratory, 2575 Sand Hill Road, Menlo Park 94025, CA, USA.\\
  $^2$Institute for Astronomy, 2680 Woodlawn Drive, Honolulu, HI 96822, USA}

\begin{document}
\date{Accepted ???, Received ???; in original form \today}
\pagerange{\pageref{firstpage}--\pageref{lastpage}} \pubyear{2008}

\maketitle
\label{firstpage}

\begin{abstract}
  We use measurements of the growth of cosmic structure, as inferred
  from the observed evolution of the X-ray luminosity function (XLF)
  of galaxy clusters, to constrain departures from General Relativity
  (GR) on cosmological scales. We employ the popular growth rate
  parameterization, $\Omega_{\rm m}(z)^{\gamma}$, for which GR
  predicts a growth index $\gamma\sim 0.55$. We use observations of
  the cosmic microwave background (CMB), type Ia supernovae (SNIa),
  and X-ray cluster gas-mass fractions ($f_{\rm gas}$), to
  simultaneously constrain the expansion history and energy content of
  the Universe, as described by the background model parameters:
  $\Omega_{\rm m}$, $w$, and $\Omega_{\rm k}$, i.e., the mean matter
  density, the dark energy equation of state parameter, and the mean
  curvature, respectively. Using conservative allowances for
  systematic uncertainties, in particular for the evolution of the
  mass--luminosity scaling relation in the XLF analysis, we find
  $\gamma=0.51^{+0.16}_{-0.15}$ and $\Omega_{\rm m}=0.27\pm 0.02$
  (68.3 per cent confidence limits), for a flat cosmological constant
  ($\Lambda$CDM) background model. Allowing $w$ to be a free
  parameter, we find $\gamma=0.44^{+0.17}_{-0.15}$. Relaxing the
  flatness prior in the $\Lambda$CDM model, we obtain
  $\gamma=0.51^{+0.19}_{-0.16}$. When in addition to the XLF data we
  use the CMB data to constrain $\gamma$ through the ISW effect, we
  obtain a combined constraint of $\gamma=0.45^{+0.14}_{-0.12}$ for
  the flat $\Lambda$CDM model. Our analysis provides the tightest
  constraints to date on the growth index. We find no evidence for
  departures from General Relativity on cosmological scales.
\end{abstract}


\section{Introduction}
\label{introduction}

Recently, using the observed evolution of the X-ray luminosity
function (XLF) of massive galaxy clusters, \citet[][hereafter
M08]{Mantz:08} presented new constraints on dark energy from
measurements of the growth of cosmic structure \cite[see
also][]{Vikhlinin:09}. These results are consistent with and
complementary to those based on measurements of the cosmic expansion
history, as deduced from distances to type Ia supernovae
\citep{Riess:98,Perlmutter:98, Davis:07, Kowalski:08}, the gas mass
fraction in galaxy clusters \citep{Allen:04, Allen:08, Rapetti:05,
  Rapetti:07}, and Baryon Acoustic Oscillations in the distribution of
galaxies \citep{Cole:05,Eisenstein:05,Percival:07}. In combination
with independent measurements of the cosmic microwave background
\citep{Spergel:03,Spergel:06,Komatsu:09}, such measurements argue for
a universe that is spatially flat, with most matter being cold and
dark, and with the energy density being currently dominated by a
cosmological constant, $\Lambda$.

Due to severe theoretical problems associated with the cosmological
constant, a plethora of other dark energy models have been proposed
\citep[for a review, see][]{Frieman:08}. Recent works \citep[][and
references therein]{Carroll:06, Caldwell:07, Hu:07, Amin:08,
  Bertschinger:08} have studied the possibility that late--time cosmic
acceleration is not driven by dark energy but rather by gravity. These
authors propose various frameworks to test for both time and
scale-dependent modifications to GR at late times and on large
scales. Although, currently, modifications to GR may be argued to be
theoretically disfavoured, it is essential to test for such deviations
using powerful, new data that are now becoming available. To measure
departures from GR on cosmological scales, experiments sensitive to
the dynamical effects of GR on these scales are required. Such
experiments include, for example, the cross-correlation of the
integrated Sachs-Wolfe (ISW) effect with other matter tracers; galaxy
clustering; weak gravitational lensing; or cluster number counts
using, e.g., X-ray selected samples.

General Relativity has been thoroughly tested from laboratory to Solar
System scales. However, GR has only just begun to be tested on
cosmological scales. Several authors have recently investigated a
simple parameterization of the growth rate, $\Omega_{\rm m}(z)^{\rm
  \gamma}$ \cite[first introduced by][]{Peebles:80}, to test for
time--dependent modifications to GR \citep[see e.g.][]{Linder:05,
  Sapone:07, Polarski:07, Gannouji:08, Acquaviva:08,
  Ballesteros:08,Mortonson:09, Thomas:09}. For a growth index,
$\gamma$, of approximately $0.55$, this parameterization accurately
models the growth rate of GR. Some authors
\citep{Nesseris:07,DiPorto:07,Wei:08,Gong:08} have recently estimated
constraints on $\gamma$ by combining results from measurements of
redshift space distortions and evolution in the galaxy power spectrum,
as well as measurements of the normalization of the matter power
spectrum, $\sigma_{8}(z)$, from Lyman-$\alpha$ forest data. Using
current cosmic shear and galaxy clustering data at low redshift,
\cite{Dore:07} placed constraints on scale-dependent modifications to
GR. These authors constrained two phenomenological, although
physically motivated, modifications of the Poisson equation on
megaparsec scales (from $0.04$ to $10\Mpc$).

In this paper, we use the XLF experiment developed by M08, and data
from the {\it ROSAT} Brightest Cluster Sample
\cite[BCS;][]{Ebeling:98}, the {\it ROSAT}-ESO Flux Limited X-ray
cluster sample \cite[REFLEX;][]{Bohringer:04}, the MAssive Cluster
Survey \cite[MACS;][]{Ebeling:01} and the 400 square degrees {\it
  ROSAT} PSPC cluster survey \cite[400sd;][]{Burenin:07}, to constrain
departures from GR on scales of tens of megaparsecs over the redshift
range $z\lt 0.9$. We use CMB \citep{Komatsu:09}, SNIa
\citep{Kowalski:08}, and cluster $f_{\rm gas}$ data \citep{Allen:08}
to simultaneously constrain the background evolution of the
Universe. We examine three background models: flat $\Lambda$CDM, flat
$w$CDM and non-flat $\Lambda$CDM. We employ a Markov Chain Monte Carlo
(MCMC) analysis, accounting for systematic uncertainties in the
experiments. Our results represent the first constraints on $\gamma$
from the observed growth of cosmic structure in galaxy clusters.

\section{Parameterizing the growth of cosmic structure}
\label{sec:gi}

In GR, the evolution of the linear matter density contrast
$\delta\equiv \delta\rho_{\rm m}/\rho_{\rm m}$, where $\rho_{\rm m}$
is the mean comoving matter density and $\delta\rho_{\rm m}$ a matter
density fluctuation, can be calculated in the synchronous gauge by
solving the scale-independent equation

\begin{equation}
  \ddot{\delta}+2\frac{\dot{a}}{a}\dot{\delta}=4G\pi\rho_{\rm m}\delta\,,
\label{eq:delta}
\end{equation}

\noindent where `dot' represents a derivative with respect to time,
and $a$ is the cosmic scale factor.

Following \cite{Lahav:91} and \cite{Wang:98}, several authors
\cite[see e.g.][]{Huterer:06, Linder:07} have parameterized the
evolution of the growth rate as $f(a)\equiv d\ln\delta/d\ln
a=\Omega_{\rm m} (a)^{\rm \gamma}$. Recasting this expression we have
the differential equation

\begin{equation}
  \frac{d\delta}{da}=\frac{\Omega_{\rm m} (a)^{\rm \gamma}}{a}\,\delta\,,
\label{eq:param}
\end{equation}

\noindent where $\gamma$ is the growth index, and $\Omega_{\rm
  m}(a)=\Omega_{\rm m} a^{-3}/E(a)^{2}$. Here, $E(a)=H(a)/H_0$ is the
evolution parameter, $H(a)$ the Hubble parameter, and $H_{\rm 0}$ its
present--day value. It has been shown \cite[see e.g.][]{Linder:07}
that for $\gamma\sim 0.55$, Equation~\ref{eq:param} accurately
reproduces the evolution of $\delta$ obtained from
Equation~\ref{eq:delta}. Using the Einstein-Boltzmann code {\sc
  camb}\footnote{http://www.camb.info/} \citep{Lewis:00}, we find that
the linear growth $\delta(a)$ obtained from Equation~\ref{eq:param}
with $\gamma\sim 0.55$ is accurate to better than $0.1$ per cent for
the relevant scales, redshifts and values of cosmological
parameters. Therefore, we adopt $\gamma\sim 0.55$ as a reference, from
which to determine departures from GR.

Equation~\ref{eq:param} provides a phenomenological model for the
growth of density perturbations that allows us to test departures from
GR without adopting a particular, fully covariant modified gravity
theory. In the absence of such an alternative gravity theory, we
perform consistency tests using convenient parameterizations of the
background expansion, within GR. We investigate three expansion models
that are well tested with current data: flat $\Lambda$CDM, a constant
dark energy equation of state, $w$CDM\footnote{This model is only used
  as a expansion model, and does not assume the presence of dark
  energy. Therefore, we do not include dark energy density
  perturbations. The evolution of the density perturbations, due only
  to matter, are modelled using $\gamma$.}, and non-flat
$\Lambda$CDM. We can write a general evolution parameter for these
models as~\footnote{Although massless neutrinos and photons are
  included in the analysis, they are neglible at late times.}

\begin{equation}
  E(a)=\left[\Omega_{\rm m} \,a^{-3}+\Omega_{\rm de} \,a^{-3(1+w)}+\Omega_{\rm k}\,a^{-2}\right]^{1/2}\,,
\label{eq:wc}
\end{equation}

\noindent where $w=-1$ for the $\Lambda$CDM models, $\Omega_{\rm de}$
is the cosmological constant/dark energy density, and $\Omega_{\rm k}$
is the curvature energy density, which is $0$ for flat models.

The growth rate $\Omega_{\rm m} (a)^{\rm \gamma}$ conveniently tends
to $1$ in the matter dominated era (high $z$), thereby matching GR for
any value of $\gamma$. Thus, we naturally match the initial value of
$\delta$ in Equation~\ref{eq:param} at high-$z$ with that of GR (see
Section~\ref{sec:xlf}).

\section{Analysis of the X-ray luminosity function}
\label{sec:xlf}

We have incorporated the growth index parameterization into the code
developed by M08. Briefly, in the XLF analysis, we compare X-ray
flux-redshift data from the cluster samples to theoretical
predictions. The relation between cluster mass and observed X-ray
luminosity is calibrated using deeper pointed X-ray observations
\citep{Reiprich:02}.

\subsection{Linear theory}
\label{sec:linear}

The variance of the linearly evolved density field, smoothed by a
spherical top-hat window of comoving radius $R$, enclosing a mass
$M=4\pi {\rho}_{\rm m} R^3 / 3$, is

\begin{equation}
  \sigma^2(M,z) = \frac{1}{2\pi^2} \int_0^\infty k^2 P(k,z) |W_{\rm M}(k)|^2 dk\,,
\label{eq:var}
\end{equation}

\noindent where $W_{\rm M}(k)$ is the Fourier transform of the window
function, and $P(k,z)\propto k^{n_{\rm s}} T^2(k,z_{\rm t})\,D(z)^2$
is the linear matter power spectrum as a function of the wavenumber,
$k$, and redshift, $z$.
Here, $n_{\rm s}$ is the scalar spectral index of the primordial
fluctuations, $T(k,z_{\rm t})$ is the matter transfer function at
redshift $z_{\rm t}$, and $D(z)\equiv\delta(z)/\delta(z_{\rm
    t})=\sigma(M,z)/\sigma(M,z_{\rm t})$
is the growth factor of linear perturbations, normalized to
unity at redshift $z_{\rm t}$. We choose $z_{\rm t}=30$, well within
the matter dominated era \citep{Bertschinger:08}. Using~\CAMB, we
calculate $T(k,z_{\rm t})$ assuming that GR is valid at early times
($z>z_{\rm t}$) and that modifications to GR are scale-invariant. For
$z<z_{\rm t}$, we calculate $D(z)$ using Equation~\ref{eq:param}.

\subsection{Non-linear N-body simulations}
\label{sec:nonlinear}

Using large N-body simulations, \cite{Jenkins:01} and \cite{Evrard:02}
showed that the mass function of dark matter halos can be conveniently
fitted by the expression

\begin{equation}
  f(\sigma^{-1}) \equiv \frac{M}{\rho_{\rm m}} \frac{dn(M,z)}{d\ln\sigma^{-1}} = A \exp \left( -|\ln \sigma^{-1}+B|^\epsilon  \right),
\label{eq:mf}
\end{equation}

\noindent where $A=0.316$, $B=0.67$, and $\epsilon=3.82$
\citep{Jenkins:01}. These fit values were determined using a spherical
overdensity group finder, at 324 times the mean matter density. This
formula is approximately `universal' with respect to the cosmology
assumed. The universality of this formula has been tested for a wide
range of cosmologies \citep{Kuhlen:04, Lokas:04, Klypin:03,
  Linder:03b, Mainini:03, Maccio:04, Francis:09, Grossi:09}. Recently
\cite{Warren:06} and \cite{Tinker:08} have also reexamined this mass
function using a larger suite of simulations.

For linear scales we use the $\gamma$-model to test the growth rate
for consistency with GR, while for non-linear scales we assume
GR\footnote{To constrain a particular gravity model, in addition to
  the expansion history and evolution of the linear density
  perturbations, we also need the mass function which encompasses the
  non-linear effects of the model. Recently \cite{Schmidt:09a} and
  \cite{Schmidt:09} have presented halo mass functions for the $f(R)$
  and DGP \citep{Dvali:00} gravity models, respectively. The former
  work shows that for $f(R)$ models compatible with Solar System
  tests, equation~\ref{eq:mf} provides a good fit to the mass function
  over the relevant mass range within the uncertainties we allow. For
  more extreme f(R) models, or for less massive clusters (for which
  non-linear chameleon effects become important) differences
  arise. For DGP, \cite{Schmidt:09} shows that the mass function for
  massive halos is significantly suppressed with respect to GR. Such
  differences can, in principle, be constrained strongly with XLF
  data.}. As in M08, we use Equation~\ref{eq:mf} to predict the number
density of galaxy clusters, $n$, at a given $M$ and $z$.


\subsection{Mass--luminosity relation}
\label{sec:ml}

Following M08, we employ a power-law mass--luminosity relation,
assuming self--similar evolution between the mass and X-ray
luminosity, $L$, of massive clusters \citep[e.g.][]{Bryan:98}. We
emphasize, however, that we include a generous allowance for
departures from self--similarity, encoded in the parameter $\zeta$,

\begin{equation}
  E(z)M_\Delta = M_0 \left[\frac{L}{E(z)}\right]^\beta (1+z)^{\zeta}\,.
  \label{eq:MLpowlaw}
\end{equation}

\noindent Here $M_{\rm\Delta}$ is the cluster mass defined at an
overdensity of $\Delta$ with respect to the critical density, and
$\log M_0$ and $\beta$ are model parameters fitted in the MCMC
analysis.

As in M08, we assume a log-normal intrinsic scatter in luminosity for
a given mass, $\eta$, including this as a model parameter. It is
possible that this scatter may evolve with redshift; we, therefore,
parameterize the evolution in the scatter as $\eta (z) = \eta_0 \left(
  1 + \eta_{\rm z} z \right)$,
where $\eta_0$ is the intrinsic scatter today, and $\eta_{\rm z}$ is a
parameter that allows for linear evolution in the scatter. Since
current data are not able to measure this evolution \citep[see
e.g.][]{Ohara:06,OHara:07,Chen:07}, we employ the same conservative
prior on $\eta_{\rm z}$ used in M08 (see below).

\subsection{Priors and systematic allowances}
\label{sec:su}

As in M08, our XLF analysis includes five parameters ($A$, $\zeta$,
$\eta_{\rm z}$, $B$, and $s_{\rm b}$)\footnote{$B$ and $s_{\rm b}$ are
  defined in Section 3.1.1 of M08. They parameterize the bias, and the
  scatter in the bias, expected for hydrostatic mass measurements from
  X-ray data.
} that are not constrained by current data. We apply the same
conservative priors on these parameters as M08.

We apply a Gaussian prior on $A$, with a mean value of $0.316$, and
standard deviation of 20 per cent. This deviation conservatively spans
the theoretical uncertainty in the mass function of
Equation~\ref{eq:mf}. For $\zeta$, we use the uniform prior
$(-0.35,0.35)$, i.e. we allow up to $\sim 25$ per cent change with
respect to self--similar evolution out to $z=1$. (We also investigate
results assuming strictly self--similar evolution,
i.e. $\zeta=0$). For $\eta_{\rm z}$, we use the uniform prior
$(-0.3,0.3)$, i.e., we allow up to $\sim 30$ per cent evolution in the
scatter to $z=1$.

\begin{figure}
\begin{center}
\includegraphics[width=3.3in]{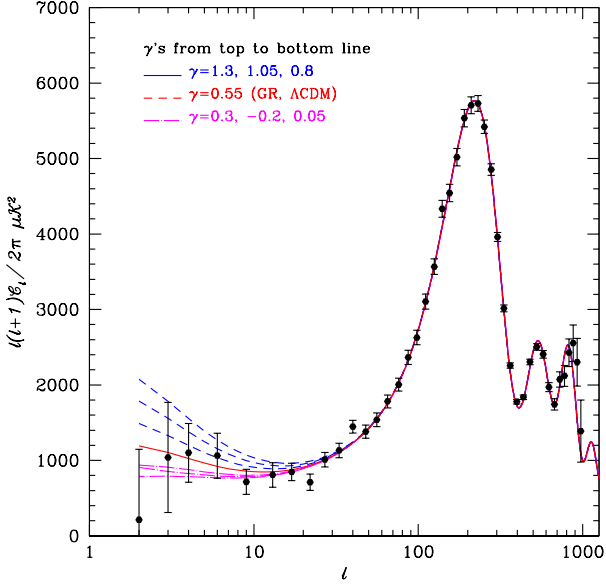}
\caption{CMB temperature anisotropy power spectra, $C_l$, for
  $\gamma=0.55$ (GR; best-fit $\Lambda$CDM model from the five-year
  WMAP data) (solid, red line), and for values of $\gamma$ higher than
  GR (dashed, blue lines), and lower (dot-dashed, magenta lines). The
  lines are equally spaced in $\gamma$. The data are more constraining
  for the higher values of $\gamma$, which produce larger changes in
  the ISW effect and significantly worse fits to the data. Higher
  values of $\gamma$ monotonically increase the ISW effect while lower
  values first decrease it and then increase it (see a similar
  behaviour in Hu 2008). The circles are the binned WMAP5 data.}
\label{fig:isw}
\end{center}
\end{figure}

Departures from hydrostatic equilibrium and sphericity introduce
well-known biases in mass measurements from X-ray data. Following M08,
we assume a $25$ per cent mean bias, and $16$ per cent scatter in the
bias, with 20 per cent systematic uncertainties in these values, as
indicated by hydrodynamical simulations and weak lensing data.

\section{Integrated Sachs-Wolfe effect}
\label{sec:isw}

Through the integrated Sachs-Wolfe (ISW) effect, the low multipoles of
the temperature anisotropy power spectrum of the CMB are sensitive to
the growth of cosmic structure, and therefore to dark energy and
modified gravity models \cite[see e.g.][]{Fang:08}. The ISW effect
arises when the gravitational potentials of large scale structures
vary with time, yielding a net energetic contribution to the CMB
photons crossing them.

On large scales, we calculate the transfer function due to the ISW
effect as \citep{Weller:03}

\begin{equation}
  \Delta_{l}^{\rm ISW}(k) = 2 \int dt\, {\rm e}^{-\tau(t)}
  \phi'j_{l}\left[k(t-t_{\rm 0})\right]
  \label{eq:isw}
\end{equation}

\noindent where `prime' denotes a derivative with respect to the
conformal time $t$, $t_{\rm 0}$ is the conformal time today, $\tau$ is
the optical depth to the last scattering, and $j_{l}(x)$ is the
spherical Bessel function for the multipole $\ell$. The total transfer
function is the sum of Equation~\ref{eq:isw} and the transfer function
from the last scattering surface (LSS), $\Delta_{l}(k)=\Delta_{l}^{\rm
  ISW}(k)+\Delta_{l}^{\rm LSS}(k)$. From this, we calculate the
anisotropy power spectrum of the temperature fluctuations $C_{l}$ (see
Figure~\ref{fig:isw}).

For GR, we can calculate the gravitational potential $\phi$ using the
Poisson equation $k^2\phi=-4 \pi G a^2\,\delta\rho_{\rm m}$, for which
$\phi$ depends only on the matter density perturbations
$\delta\rho_{\rm m}$\footnote{Using {\sc camb} we have verified this
  for the relevant region of parameter space. The other two components
  of the Weyl tensor that would contribute to $\phi$, the anisotropic
  stress and energy flux \cite[for details on these terms
  see][]{Challinor:98}, are negligible.}. Here we use the
$\gamma$-model to test CMB data for consistency with GR, restricting
ourselves to departures from GR in which the ISW effect can be
calculated, as in GR, using the Poisson equation at all
scales\footnote{For a general gravity model, none of the components of
  the Weyl tensor will be negligible and, therefore modifications to
  the Poisson equation will be required \cite[see e.g.][]{Hu:07,
    Hu:08}. Interestingly, for the DGP gravity model,
  \cite{Sawicki:06} showed that the used of an appropiately modified
  Poisson equation near the horizon yields an enhancement of the ISW
  effect, and thus stronger constraints with CMB data.}.

The ISW effect is only relevant at $z\lt 2$. For this redshift range,
we modify {\sc camb} to calculate the evolution of $\delta$, and
therefore $\phi'$, using Equation~\ref{eq:param}, and obtain

\begin{equation}
  \phi' = 4 \pi G\,\frac{a^2}{k^2}\,\dot{a}\,\delta\rho_{\rm m}\left[1-\Omega_{\rm m}(a)^{\gamma}\right]\,.
\label{eqn:ISW}
\end{equation}

\noindent We calculate $\Delta_{l}^{\rm ISW}(k)$ at $z=2$ using GR,
and use these values as initial conditions when evolving to $z=0$ in
the $\gamma$-model. 

\begin{figure*}
\begin{center}
\includegraphics[width=2.3in]{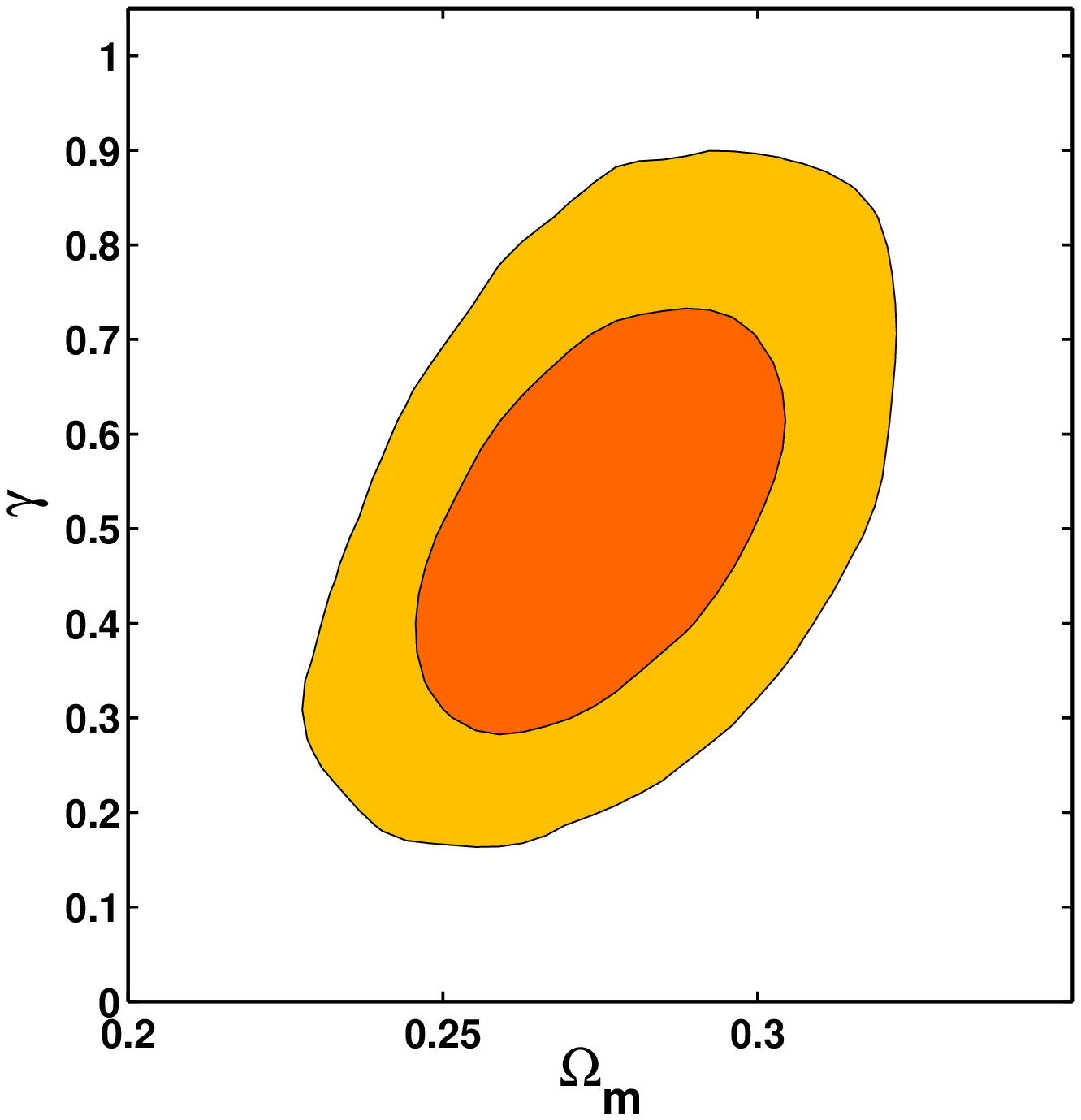}
\hspace{1.1cm}
\includegraphics[width=2.3in]{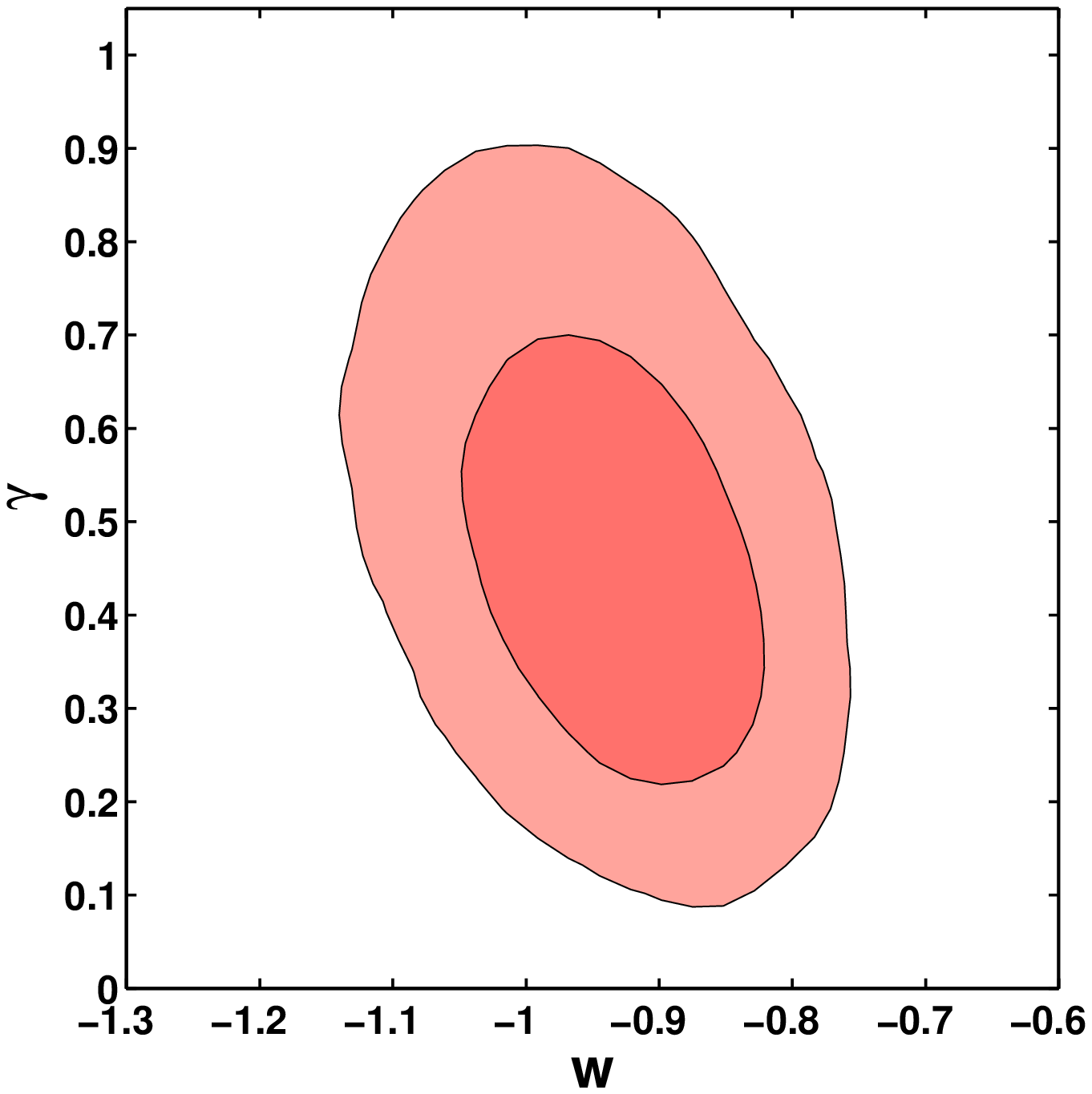}
\caption{68 and 95 per cent confidence contours in the (left panel)
  $\Omega_{\rm m},\gamma$ plane for the flat $\Lambda$CDM background
  model, and (right panel) $w,\gamma$ plane for the flat $w$CDM model,
  from the combination of XLF, CMB, SNIa, and $f_{\rm gas}$
  data. These results do not include the (subdominant) contribution
  from the ISW effect (see text and Table~\ref{table:results} for
  details).}
\label{fig:models}
\end{center}
\end{figure*}

\section{Data analysis}
\label{sec:data}

We use the five-year WMAP CMB data \cite[][and references
therein]{Dunkley:09, Komatsu:09}, SNIa data from the ``Union''
compilation of \cite{Kowalski:08}, the $f_{\rm gas}$ measurements of
\cite{Allen:08}, and XLF data from the BCS~\citep[$z\lt 0.3$, northern
sky;][]{Ebeling:98}, REFLEX~\citep[$z\lt 0.3$, southern
sky;][]{Bohringer:04}, MACS~\citep[$0.3\lt z\lt 0.7$, $\sim 55$ per
cent sky coverage;][]{Ebeling:01, Ebeling:07}, and 400sd~\citep[$z\lt
0.9$, $\sim 1$ per cent sky coverage;][]{Burenin:07} cluster
samples. Adopting a luminosity limit of $2.55\E{44}
h_{70}^{-2}\erg\second^{-1}$ ($0.1-2.4\keV{}$), we use 78 clusters
above a flux limit of $4.4\times 10^{-12} \erg \second^{-1} \cm^{-2}$
from BCS, 130 above $3.0\times 10^{-12} \erg\second^{-1} \cm^{-2}$
from REFLEX, 34 above $2\times 10^{-12} \erg \second^{-1} \cm^{-2}$
from MACS, and 30 above $0.14\times 10^{-12} \erg \second^{-1}
\cm^{-2}$ from the 400sd survey.

We use a Metropolis MCMC algorithm to calculate the posterior
probability distributions of our model parameters. We use a modified
version of the \COSMOMC\footnote{http://cosmologist.info/cosmomc/}
code \citep{Lewis:02} that includes additional modules to calculate
the likelihood for the $f_{\rm gas}$
experiment\footnote{http://www.stanford.edu/$\sim$drapetti/fgas\_module/}
\citep{Allen:08}, and for the XLF experiment (M08).

When we use flat $\Lambda$CDM as background model we fit seven
cosmological parameters: the mean physical baryon density,
$\Omega_{\rm b}h^2$; the mean physical dark matter density,
$\Omega_{\rm c}h^2$; the (approximate) ratio of the sound horizon at
last scattering to the angular diameter distance, $\theta$ \cite[which
is less correlated with other parameters than $H_0$, as shown
by][]{Kosowsky:02}; the optical depth to reionization, $\tau$; the
adiabatic scalar spectral index, $n_{\rm s}$; the logarithm of the
adiabatic scalar amplitude, $\ln(A_{\rm s})$; and the growth index,
$\gamma$. For the flat $w$CDM background model, we have $w$ as an
additional parameter; and for the non-flat $\Lambda$CDM model,
$\Omega_{\rm k}$. We marginalize over seven other parameters that
model systematic uncertainties in the $f_{\rm gas}$ analysis \cite[see
details in][]{Allen:08}, and seven more that account for uncertainties
in the XLF analysis. For the SNIa analysis, we use, as provided by
\cite{Kowalski:08}, a covariance matrix that accounts for systematic
uncertainties.

\section{Results}
\label{sec:constraints}

As shown in \cite{Allen:08} and references therein, the combination of
$f_{\rm gas}$, SNIa, and CMB data places tight constraints on the
expansion history and $\Omega_{\rm m}$. The addition of the XLF data
allows us to place tight constraints on $\gamma$. Assuming a flat
$\Lambda$CDM background model, we obtain the results shown in the left
panel of Figure~\ref{fig:models}. The results for a flat, constant-$w$
model are shown in the right panel of Figure~\ref{fig:models}. The
marginalized values are summarized in Table~\ref{table:results}. For
flat $\Lambda$CDM we find
$\gamma=0.51^{+0.16}_{-0.15}$. Interestingly, when allowing $w$ to be
free, we obtain similar constraints of
$\gamma=0.44^{+0.17}_{-0.15}$. We also see an anticorrelation between
$\gamma$ and $w$, wherein models close to a cosmological constant
($w=-1$) are most consistent with GR ($\gamma\sim0.55$). We have also
investigated the effect of relaxing the assumption of flatness in the
$\Lambda$CDM model and find $\gamma=0.51^{+0.19}_{-0.16}$, with no
significant covariance between $\gamma$ and $\Omega_{\rm k}$.

For massive clusters, such as those in the current XLF data set,
self--similar evolution of the mass--luminosity relation is a
well--motivated theoretical prediction \citep{Bryan:98}. Current
Chandra data (Mantz et al., submitted) show good agreement with this
prediction over the redshift range spanned by the XLF data ($z\lt
0.5$). Under the assumption of strictly self--similar evolution
($\zeta=0$), we find $\gamma=0.44^{+0.12}_{-0.11}$ for the flat
$\Lambda$CDM background model.

As discussed in Section~\ref{sec:isw}, the low multipoles of the CMB
are also sensitive to the growth rate through the ISW effect. Even
though the ISW data currently have less constraining power on $\gamma$
than the XLF\footnote{Note that the cross--correlation of the ISW
  effect with galaxy surveys \citep[see e.g.][]{Ho:08, Lombriser:09}
  may provide competitive, additional constraint on $\gamma$.}, adding
the ISW leads to $\sim 17$ per cent tighter constraints on $\gamma$:
for the flat $\Lambda$CDM background model we obtain
$\gamma=0.45^{+0.14}_{-0.12}$, for flat $w$CDM
$\gamma=0.42^{+0.14}_{-0.13}$, and $\gamma=0.50^{+0.16}_{-0.15}$ for
non-flat $\Lambda$CDM.

\begin{table}
\begin{center}
  \caption{Marginalized $1\sigma$ constraints on $\Omega_{\rm m}$,
    $\sigma_8$, and $\gamma$, using all the data sets without
    (default) or with (default+ISW) the ISW effect, for the flat
    $\Lambda$CDM (A), flat $w$CDM (B), and non-flat $\Lambda$CDM (C)
    expansion history models. For the B model we obtain
    $w=-0.93\pm0.07$. The results for $\Omega_{\rm m}$ are the same
    with or without the ISW.}
\label{table:results}
\begin{tabular}{ c c c }
               & $\Omega_{\rm m}, ~\sigma_8, ~\gamma$ (default)  & $\sigma_8, ~\gamma$ (default+ISW)              \\
\hline                                                                                                                                                                                                           
\noalign{\vskip 3pt}                                                                                                                                                                                             
A    & $0.27^{+0.02}_{-0.02}, ~0.82^{+0.05}_{-0.04}, ~0.51^{+0.16}_{-0.15}$ & $0.83^{+0.04}_{-0.04}, ~0.45^{+0.14}_{-0.12}$ \\
\noalign{\vskip 4pt}                                                                                                                                                                                             
B    & $0.28^{+0.02}_{-0.02}, ~0.81^{+0.05}_{-0.05}, ~0.44^{+0.17}_{-0.15}$ & $0.82^{+0.04}_{-0.04}, ~0.42^{+0.14}_{-0.13}$ \\
\noalign{\vskip 4pt} 
C    & $0.31^{+0.04}_{-0.03}, ~0.76^{+0.06}_{-0.04}, ~0.51^{+0.19}_{-0.16}$ & $0.77^{+0.05}_{-0.05}, ~0.50^{+0.16}_{-0.15}$ \\
\end{tabular}
\end{center}
\end{table}

\section{Conclusions}
\label{sec:conclusions}

Combining XLF, $f_{\rm gas}$, SNIa and CMB data, we have
simultaneously constrained the background evolution of the Universe
and the growth of matter density fluctuations on cosmological scales,
allowing us to search for departures from GR. We parameterize the
expansion history with simple models that include late-time cosmic
acceleration, taking flat $\Lambda$CDM as our default model, but also
investigating flat, constant $w$ models, and non--flat $\Lambda$CDM
models. We parameterize the growth of cosmic structure using the
growth index, $\gamma$, which assumes the same scale dependence as GR,
but allows for time--dependent deviations from it.

We have performed MCMC analyses with seven (or eight) interesting
cosmological parameters, and an additional fourteen parameters to
encompass conservative allowances for systematic
uncertainties. Marginalizing over these allowances, we obtain the
tightest constraints to date on the growth index. For the flat
$\Lambda$CDM background model, we measure
$\gamma=0.51^{+0.16}_{-0.15}$. Allowing $w$ or $\Omega_{\rm k}$ to be
free, we obtain similar constraints: $\gamma=0.44^{+0.17}_{-0.15}$ and
$\gamma=0.51^{+0.19}_{-0.16}$, respectively. Including also
constraints on $\gamma$ from the ISW effect, we obtain
$\gamma=0.45^{+0.14}_{-0.12}$ for flat $\Lambda$CDM. Currently, we
find no evidence for departures from General Relativity.

In the near future, improved XLF analyses should provide significantly
tighter constraints on both $\gamma$ and $\zeta$, allowing us to
explore more complex modified gravity models (Rapetti et al., in
preparation). Our results highlight the importance of X-ray cluster
data, and the potential of combined expansion history plus growth of
structure studies, for testing dark energy and modified gravity models
for the acceleration of the Universe.

\section*{Acknowledgments}

We thank A. Lewis, M. Amin, R. Blandford, P. Wang, W. Hu, L. Lombriser
and I. Sawicki for useful discussions. We thank G.~Morris for
technical support. The computational analysis was carried out using
the KIPAC XOC and Orange computer clusters at SLAC. We acknowledge
support from the National Aeronautics and Space Administration through
Chandra Award Numbers DD5-6031X and G08-9118X issued by the Chandra
X-ray Observatory Center, which is operated by the Smithsonian
Astrophysical Observatory for and on behalf of the National
Aeronautics and Space Administration under contract NAS8-03060. This
work was supported in part by the U.S.  Department of Energy under
contract number DE-AC02-76SF00515.

\bibliography{biblist_growthindex}
\bibliographystyle{mn2e}

\label{lastpage}
\end{document}